\documentclass[10pt,aps,prd,twocolumn,notitlepage,nofootinbib,floatfix,superscriptaddress]{revtex4-1}
\usepackage{graphicx}  
\usepackage{amsmath}   
\usepackage{amssymb}   
\usepackage{bm} 
\usepackage{dcolumn}
\usepackage{color}
\usepackage{mathrsfs}
\usepackage{gensymb}
\usepackage{float}
\usepackage{amsfonts}
\usepackage{varioref}
\usepackage{textcomp}
\usepackage[caption=true]{subfig}
\RequirePackage[colorlinks,citecolor=blue,urlcolor=magenta,linkcolor=blue]{hyperref}
\allowdisplaybreaks

\def\gr{general relativity}

\labelformat{section}{Section #1} 
\labelformat{subsection}{Section #1} 
\labelformat{subsubsection}{Section #1}
\labelformat{subsubsubsection}{Section #1}
\labelformat{equation}{Eq.~(#1)} 
\labelformat{figure}{Fig.~#1} 
\labelformat{subfigure}{Fig.~\thefigure#1} 
\labelformat{table}{Tab.~#1} 
\labelformat{appendix}{Appendix #1}
\begin{document}

\title{\bf Hunting extra dimensions in the shadow of Sgr A*}

\author{Indrani Banerjee}
\email{banerjeein@nitrkl.ac.in}
\affiliation{Department of Physics and Astronomy, National Institute of Technology, Rourkela}

\author{Sumanta Chakraborty} 
\email{tpsc@iacs.res.in}
\affiliation{School of Physical Sciences, Indian Association for the Cultivation of Science, Kolkata-700032, India}

\author{Soumitra SenGupta}
\email{tpssg@iacs.res.in}
\affiliation{School of Physical Sciences, Indian Association for the Cultivation of Science, Kolkata-700032, India}
\begin{abstract}
We show that the observed angular diameter of the shadow of the ultra compact object Sgr A*, favours the existence of an extra spatial dimension. This holds irrespective of the nature of the ultra compact object, i.e., whether it is a wormhole or, a black hole mimicker, but with the common feature that both of them have an extra dimensional origin. This result holds true for the mass and the distance measurements of Sgr A* using both Keck and the Gravity collaborations and whether we use the observed image or, the observed shadow diameter. In particular, the central value of the observed shadow or, the observed image diameter predicts non-zero hairs inherited from the extra dimensions. 
\end{abstract}
\maketitle
\section{Introduction}

Knowledge about the nature of the gravitational interaction in the near-horizon regime is expected to lead towards a better understanding of gravity and the nature of ultra-compact objects at a fundamental level. We hope to achieve this, by using the following two observations --- (a) the detection of gravitational waves from the merger of binary compact objects \cite{Abbott:2016blz,TheLIGOScientific:2017qsa} and (b) imaging the shadow of the supermassive compact objects at the centre of the neighbouring galaxies, e.g., the M87 and the milky way \cite{Akiyama:2019cqa,Akiyama:2019brx,EventHorizonTelescope:2022urf}. Predictions from the black hole spacetimes in general relativity, match well with the observations related to the gravitational waves and the shadow \cite{Abbott:2018lct,EventHorizonTelescope:2022xqj}. However there are a few theoretical and observational reasons, which compels one to look for theories beyond \gr, these include --- (i) prediction of singularity in black hole and cosmological spacetimes \cite{Penrose:1964wq,Hawking:1976ra,Christodoulou:1991yfa}, (ii) existence of exotic matter fields to explain the galactic rotation curves as well as the late time acceleration of the universe \cite{Milgrom:1983pn,Bekenstein:1984tv,Perlmutter:1998np,Riess:1998cb}, (iii) smooth extension of the classical metric beyond the Cauchy horizon, resulting into loss of predictability \cite{Dafermos:2003wr,Cardoso:2017soq,Rahman:2018oso}, and finally, the appearance of inevitable divergences in the quantum description of general relativity \cite{HAWKING198839,Burgess:2003jk}. In this context, it seems legitimate to explore various alternatives to general relativity, and in particular the black hole paradigm and to test them against both gravitational wave and shadow measurements.

Implications of the gravitational wave measurements for various alternative theories of gravity, as well as for the non-black hole, ultra-compact objects in general relativity have been discussed in detail in recent literatures \cite{Chakravarti:2019aup,Chakravarti:2018vlt,Chakraborty:2017qve,Visinelli:2017bny}. Similar extensive discussions also exist for the shadow measurements from M87*, the supermassive ultra-compact object at the centre of the M87 galaxy. In particular, the results from our previous work \cite{Banerjee:2019nnj}, pointed towards a very interesting possibility --- the existence of a spatial extra dimension seems to be more consistent with the observed shadow of M87* --- at least within the errors mentioned by the Event Horizon collaboration \cite{Akiyama:2019cqa}. As a natural extension we further explore whether the recently observed shadow of the supermassive ultra-compact object $\textrm{Sgr~A}^*$ also favours the possibly of an extra spatial dimensions. In this context, we would like to emphasize that for the existence of a shadow, the object need not be a black hole or, an ultra compact object, it can also be an wormhole, which is also natural in the presence of extra spacetime dimension\footnote{Again, these wormhole solutions originate from the existence of an extra spatial dimensions, but most intriguingly \emph{without} the need for any exotic matter} and must be considered in an equal footing. Thus in this work, we will not only explore the signatures of extra dimensions through the shadow of ultra-compact objects, but also through possible wormhole solutions. Of course, despite the nature of the underlying object, the key motivation is to probe any possible signatures of extra dimension in the shadow of $\textrm{Sgr~A}^*$. 

The role and possibility of existence of extra spatial dimension has earlier been explored in the context of electromagentism-gravity unification, naturalness problem in standard model of particle physics, as well as in certain models of quantum gravity, which are consistent only in the presence of extra spatial dimension \cite{ArkaniHamed:1998rs,Antoniadis:1998ig,Randall:1999ee}. Our exploration 
of extra dimensions in this work, using observable imprints of the same on the strong field tests for gravitational interaction will have implications on a much broader scale. The particular model we will consider in our work, namely the braneworld model, imprints the higher dimensional contributions in the four dimensional gravitational field equations by the higher dimensional Weyl tensor \cite{Shiromizu:1999wj,Dadhich:2000am,Harko:2004ui,Carames:2012gr,Chakraborty:2016gpg,Chakraborty:2015taq}. As a consequence, the four dimensional spacetime inherits a tidal charge term, which unlike the Reissner-Nordstr\"{o}m solution in general relativity, can appear with an opposite signature \cite{Dadhich:2000am,Harko:2004ui,Chakraborty:2014xla,Aliev:2005bi}. Besides other constraints on this tidal charge term from various different observational avenues \cite{Bhattacharya:2016naa,Banerjee:2017hzw,Banerjee:2019sae,Banerjee:2019cjk,Chakravarti:2019aup,Chakravarti:2018vlt,Chakraborty:2017qve,Visinelli:2017bny,Vagnozzi:2019apd}, our work involving black hole shadow \cite{Banerjee:2019nnj} has demonstrated that having a tidal charge is more preferred than having none. In this work we will assess, whether this potentially interesting result also holds true in the context of the recent observational data from the shadow measurement of $\textrm{Sgr~A}^*$. For completeness, we will also explore the nature of the object, sourcing the gravitational interaction, but hidden beneath the shadow.

The organization of the paper is as follows: In \ref{brane_ECO} we will summarize the spacetime geometry outside the ultra-compact object on the brane and its non-trivial contribution to the black hole shadow. Then in \ref{brane_wormhole}, we present the spacetime geometry of a braneworld wormhole, and shall study the associated shadow structure. Finally, in \ref{Sec4} we will compare the various theoretical observables of the shadow with the corresponding results from the observation of the shadow of $\textrm{Sgr~A}^*$, in order to assert the consistency of the higher dimensions with the observations. We then conclude in \ref{Sec5}.  

\section{Ultra compact object on the brane and its shadow}\label{brane_ECO}

In this section, we will present the vacuum spacetime geometry outside the ultra-compact object on the four dimensional spacetime, inheriting properties from the extra spatial dimensions. In particular, this four dimensional spacetime is an exact solution of the effective gravitational field equation on the brane (short form for the four-dimensional spacetime we live in), which can be obtained by projecting the bulk (short form for the higher dimensional spacetime) Einstein's equations on the brane. The projection involves use of the Gauss-Codazzi equation and its variant, such that we obtain the four dimensional Einstein tensor, projected from the five dimensional one. Such a projection also brings along with it the extrinsic curvatures on the brane hypersurface, with the brane being embedded in a bulk spacetime with a negative cosmological constant. The brane, on the other hand, is free from this cosmological constant, as the tensions on the brane cancels the negative cosmological constant through junction conditions, thereby fixing the extrinsic curvature terms. Finally, the effective field equations on the vacuum brane takes the form \cite{Shiromizu:1999wj,Dadhich:2000am,Harko:2004ui,Chakraborty:2015bja},
\begin{align}\label{effeq}
~^{(4)}G_{\mu \nu}+E_{\mu \nu}=0~.
\end{align}
The right hand side of the above equation vanishes, since we are interested in the case of vacuum brane. Here, as evident, $\,^{(4)}G_{\mu \nu}$ is the Einstein tensor constructed out of the brane geometry alone and $E_{\mu \nu}=W_{PAQB}e^{P}_{\mu}n^{A}e^{Q}_{\nu}n^{B}$, represents the non-local effects of the bulk through the bulk Weyl tensor $W_{ABCD}$ and its appropriate projections through the projectors $e^A_\alpha$ and the normalized normals to the brane $n^C$ \cite{Shiromizu:1999wj,Dadhich:2000am,Germani:2001du,Maartens:2003tw}. The above equation can be solved exactly, by noting that the tensor $E_{\mu \nu}$ is traceless and divergencefree, hence behaves very much like the Maxwell stress-tensor, such that we obtain the spacetime geometry as,
\begin{widetext}
\begin{align}\label{papereq1}
ds^{2}&=-\bigg(\frac{\Delta-a^{2}\sin^{2}\theta}{\rho^{2}}\bigg)dt^2-\frac{2a\sin^2\theta\left(r^{2}+a^{2}-\Delta\right)}{\rho^{2}} dt d\phi 
+ \frac{\rho^{2}}{\Delta}dr^2+\rho^{2} d\theta^2 
\nonumber
\\
&\hskip 4 cm +\bigg\{r^2 + a^2 +\frac{a^2 \sin^2\theta\left(r^{2}+a^{2}-\Delta\right)}{\rho^{2}}\bigg\}\sin^2\theta d\phi^2~,
\end{align}
\end{widetext}
where, $\rho^{2}\equiv r^2+a^2 \cos^2\theta$ and $\Delta \equiv r^2 -2Mr+a^2+M^{2}q$ \cite{Aliev:2009cg}, whose largest zero is located at $r_{+}=M+\sqrt{M^{2}-a^{2}-M^{2}q}$. Here $M$ corresponds to the mass and $J=aM$ is the angular momentum of the central compact object, with $q$ being the dimensionless tidal charge parameter inherited from the higher dimensions. Note that, the charge parameter $q$ can assume both positive and negative values and thus provides a true signature of the additional spatial dimensions \cite{Aliev:2005bi,Aliev:2009cg,Neves:2012it}. Moreover, the negative values of $q$ actually supports the rotation parameter to be larger than unity, since the existence of a $r_{+}$ demands, $(a/M)^{2}\leq 1+|q|>1$, which is another tantalizing signature for the rotating braneworld black hole.

Even though the above spacetime very much looks like that of a black hole with a horizon located at $r_{+}$, the largest root of the algebraic equation $\Delta=0$, but as we have argued in \cite{Dey:2020lhq,Dey:2020pth}, it is better to consider this object as an ultra-compact object \emph{without} any exotic matter. The reason being twofold, first of all the horizon at $r_{+}$ is not an event horizon, rather an apparent horizon, since the above spacetime cannot be extended fully to the global five dimensional spacetime \cite{Chamblin:2000ra} and for an apparent horizon there can be propagation out of the horizon as well. Secondly, these solutions are often affected by putative quantum effects near the apparent horizon $r_{+}$. This is because, the five dimensional bulk within which the brane is embedded is anti-de Sitter (AdS) in nature, owing to the existence of a negative bulk cosmological constant. Thus, assuming the AdS/CFT correspondence to be correct, it follows that there will be a quantum conformal field theory (CFT) on the brane. Hence the effective field equations, as in \ref{effeq}, should have a non-zero right hand side, depending on the expectation values of the CFT stress tensor, resulting into quantum corrections to the horizon. As a consequence of these, the horizon of the solution on the brane is not a one-way membrane. Thus we will interpret this solution as an ultra-compact object with radius close to that of $r_{+}$ and with a reflectivity, both being determined by possible quantum effects near the horizon.   

\begin{figure}
\subfloat[The shadow for the ultra-compact object on the brane has been presented with $(a/M)=0.7$ and with an inclination angle $30^{\circ}$, for various choices of the tidal charge parameter $q$.]
{\includegraphics[scale=0.45]{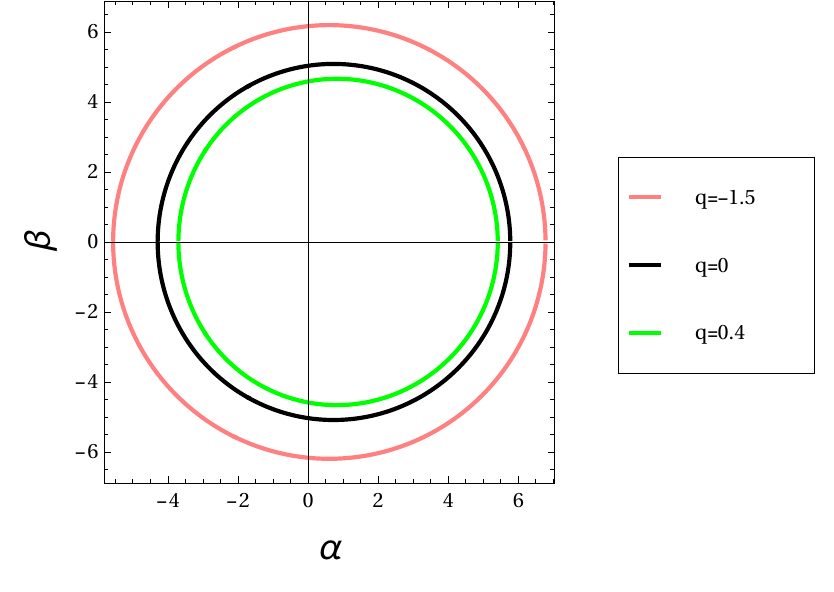}}
\hskip 1 cm 
\subfloat[The above figure demonstrates the shadow for both positive and negative values of the tidal charge parameter for an ultra-compact object in the braneworld, with $(a/M)=0.7$ and inclination angle $75^{\circ}$.]
{\includegraphics[scale=0.45]{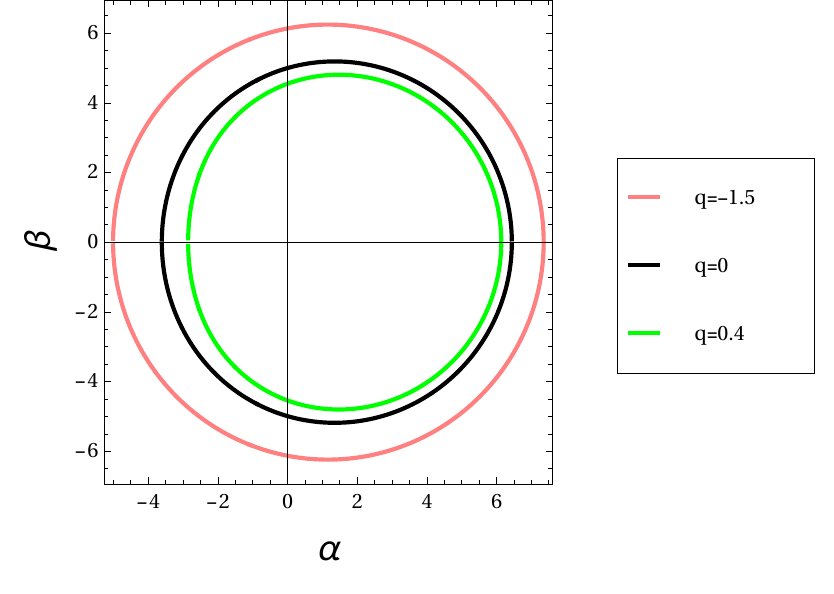}}
\caption{Shadow structures for an ultra-compact object on the brane have been plotted in the plane of celestial coordinates $(\alpha,\beta)$ for various choices of the black hole hairs and inclination angle.}
\label{Fig1}
\end{figure}

The details regarding the black hole shadow associated with the above spacetime can be found in \cite{Banerjee:2019nnj}, here we simply quote some of the results for completeness. The starting point of the computation regarding black hole shadow starts from the geodesic equations of a photon, which are separable in the above spacetime, due to the presence of a Killing tensor. From the three constants of motion, the energy $E$, angular momentum $L$ and the Carter constant $K$ \cite{Carter:1968rr}, one may introduce two impact parameters $\xi\equiv (L/E)$ and $\eta\equiv(K/E^{2})$, these denote the distance of the photon from the axis of rotation and the equatorial plane, respectively. Finally, the construction of the shadow in the observer's sky, with the observer located at a large distance $r_{0}$ from the ultra-compact object, with an inclination angle $\theta_{0}$ from the rotation axis, requires defining two celestial coordinates $\alpha$ and $\beta$, such that \cite{Bardeen:1973tla,Vries_1999,Cunha:2018acu},
\begin{align}
\alpha&=\lim_{r_{0}\rightarrow \infty}\left(-r_{0}^{2}\sin \theta_{0}\frac{d\phi}{dr}\right)=-\xi~ \textrm{cosec}\theta_{0}~;
\label{eqalpha}
\\
\beta&=\lim_{r_{0}\rightarrow \infty}\left(r_{0}^{2}\frac{d\theta}{dr}\right)=\pm \sqrt{\eta+a^{2}\cos^{2}\theta_{0}-\xi^{2}\cot^{2}\theta_{0}}~.
\label{eqbeta}
\end{align}
From which one obtains the contour of the shadow in the observer's sky, dependent on the inclination angle $\theta_{0}$. The shape and size of the shadow depends primarily on the geometry of the background spacetime and in this case on the tidal charge parameter $q$. In particular, as \ref{Fig1} demonstrates, the presence of a negative tidal charge parameter enhances the size of the shadow and that of a positive tidal charge parameter decreases the size of the shadow, when compared to \gr. To illustrate the dependence of the black hole shadow on the black hole hairs, we present in \ref{Fig1} the variation of the shadow with tidal charge $q$ and the inclination angle $\theta_0$. We will use these features to compute the observables which in turn will be compared with the observed shadow of $\textrm{Sgr~A}^*$. 

\section{Wormhole on the brane and its shadow}\label{brane_wormhole}

In an identical set up one can arrive at another class of compact object on the brane, namely wormhole. Even though wormholes, connecting two different universes, generically require exotic matter field at the throat in order for it to be traversable \cite{Bronnikov:1973,Morris:1988cz,Damour:2007ap,Bueno:2017hyj,Biswas:2022wah}, that is not the case on the braneworld scenario \cite{Kar:2015lma,Bronnikov:2002,Bronnikov:2003,Dadhich:2001fu}. In particular, the contribution from the extra spatial dimension actually behaves as an exotic matter field, so that the traversable nature of the wormhole spacetime is preserved, even when matter field on the brane satisfies all the energy conditions. In this case, rather than solving for the bulk spacetime, an expression for $E_{\mu \nu}$ can be obtained by expanding the same in the ratio of the (bulk/brane) curvature length scale. Such an expansion in a two brane system, located at $y=0$ (Planck brane) and at $y=l$ (visible brane), respectively, yields the following effective gravitational field equation on the visible brane \cite{Kar:2015lma},
\begin{widetext}
\begin{align}\label{EE}
G_{\mu \nu}&=\frac{{\bar\kappa}^2}{l\Phi} T^{\rm vis}_{\mu \nu} + \frac{{\bar\kappa}^2\,(1 + \Phi)}{l\Phi} 
T^{\rm Pl}_{\mu \nu} + \frac{1}{\Phi}\left ({\nabla}_{\mu} {\nabla}_{\nu} \Phi - g_{\mu \nu} {\nabla}^{\alpha} {\nabla}_{\alpha} \Phi\right )  
\nonumber
\\
&\hskip 2 cm -\frac{3}{2\Phi(1+\Phi)} \left ({\nabla}_{\mu}\Phi{\nabla}_{\nu}\Phi
-\frac{1}{2} g_{\mu \nu} {\nabla}^{\alpha}\Phi{\nabla}_{\alpha} \Phi\right)~.
\end{align}
\end{widetext}
Here, $g_{\mu\nu}$ is the metric on the visible brane and the covariant differentiation is defined with respect to the same metric. 
Moreover, ${\bar\kappa}^{2}$ is the five dimensional gravitational coupling constant, $T^{\rm vis}_{\mu \nu}$ and $T^{\rm Pl}_{\mu \nu}$ are the stress-energy tensors on the Planck brane and the visible brane, respectively. The scalar field $\Phi$ is known as the radion field and it measures the inter-brane separation and satisfies the following evolution equation \cite{Kar:2015lma},
\begin{equation}
{\nabla}^{\alpha}{\nabla}_{\alpha}\Phi =\frac{{\bar\kappa}^2}{l}\frac{T^{\rm vis} + T^{\rm Pl}}{2\omega+3} - 
\frac{1}{2\omega +3} \frac{d\omega}{d\Phi} ({\nabla}^{\alpha}\Phi)({\nabla}_{\alpha}\Phi)~,
\end{equation}
with $T^{\rm vis}$, $T^{\rm Pl}$ being the traces of energy momentum tensors on the Planck and the visible branes, respectively.
The coupling function $\omega({\Phi})$, appearing in the above field equation resembles the above set of equations to be that of Brans-Dicke gravity, and can be written in terms of $\Phi$ as, 
\begin{equation}
\omega (\Phi) = -\frac{3\Phi}{2(1+\Phi)}
\end{equation}
Therefore, even though both $T^{\rm vis}_{\mu \nu}$ and $T^{\rm Pl}_{\mu \nu}$ can satisfy energy conditions, violation of the same for the part arising from the radion field $\Phi$ can lead to an overall violation of the energy conditions. This is what will lead to the wormhole without exotic matter. 

The field equations, presented in \ref{EE} can be solved by assuming, $T_{\mu \nu}^{\rm Pl}=0$ and choosing $T_{\mu \nu}^{\rm vis}$ to represent anisotropic perfect fluid, such that the trace $T^{\rm vis}$ vanishes identically. This leads to the metric of the braneworld wormhole as, 
\begin{align}
ds^2&=-\left (\kappa +\lambda \sqrt{1-\frac{2M}{r}}\right )^2 dt^2 
\nonumber
\\
&\hskip 1 cm +\left(1-\frac{2M}{r}\right)^{-1}dr^2+r^2 d\Omega_{2}^{2}~,
\end{align}
It is clear that there are no curvature singularities in this line element and also the null surface at $r=2M$ (where, $g^{rr}$ vanishes) does not coincide with the infinite redshift surface (as, $g_{00}$ does not vanish at $r=2M$), as long as the parameters $\kappa$ and $\lambda$ are both positive (negative) and non-zero, with $\vert \kappa\vert > \vert \lambda \vert$. Therefore, $r=2M$ is not the Killing horizon, rather represents the location of the wormhole throat. The radion field, on the other hand, has the following solution, 
\begin{equation}
\sqrt{1+\Phi(r')}=\frac{C_{1}}{M\lambda}\ln \left[\frac{2qr'+M}{2r'+M}\right]+C_{4}~,
\end{equation}
where, $r'$ denotes the isotropic coordinate, related to the spherically symmetric coordinate $r$, through the relation, 
$r= r'[1+(M/2r')]^2$ and $q=[(\kappa+\lambda)/(\kappa-\lambda)]$. Note that for $\kappa=0$ and $\lambda=1$, the scalar field becomes a simple constant and the spacetime metric reduces to that of the Schwarzschild spacetime. Thus the presence of $\kappa$ is essential for the wormhole nature of the spacetime, having its origin in the higher dimensional physics, namely the radion field. In this work, we will look for possible signatures of non-zero values of $\kappa$ in the shadow measurements of the $\textrm{Sgr~A}^*$.

\begin{figure}
\begin{center}
\includegraphics[scale=0.45]{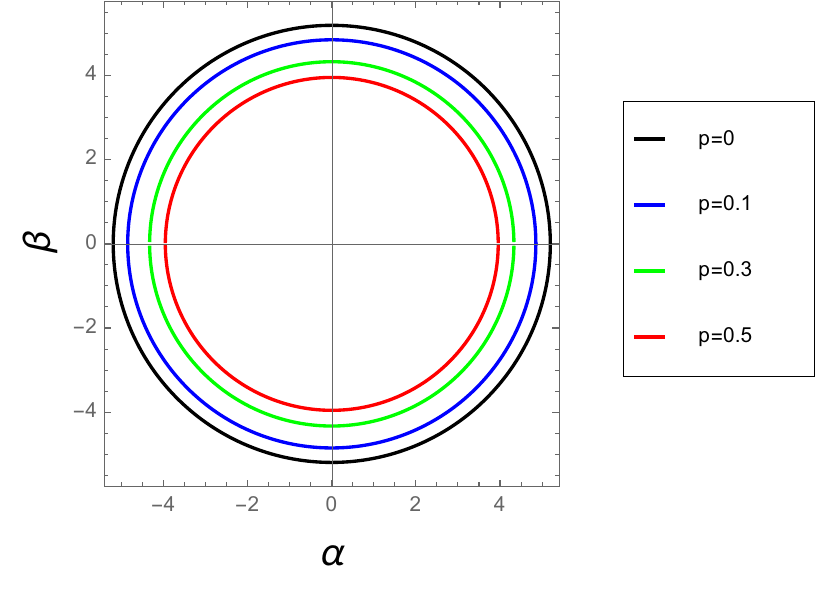}
\caption{Shadow structure for the braneworld wormhole has been plotted in the plane of the celestial coordinates $(\alpha,\beta)$ for various choices of the ratio $p=(\kappa/\lambda)$. The wormhole spacetime being static and spherically symmetric, the shadow is circular, irrespective of the choice of the inclination angle.}
\label{Fig1-1}
\end{center}
\end{figure}

It is to be emphasized that the wormhole solution considered here has not been explored earlier in the context of wormhole physics. In particular, this differs considerably from the wormhole solutions considered in \cite{Vagnozzi:2022moj} and also from those in \cite{EventHorizonTelescope:2022xqj}. For example, in \cite{Vagnozzi:2022moj}, the Damour-Solodukhin wormhole solution is considered, which is very different from the solution considered here, in their origin and also the basic physical properties, see \cite{Biswas:2022wah}. The same being true for the wormhole model considered in \cite{EventHorizonTelescope:2022xqj} as well. Thus, here we present the first attempt to describe the shadow of $\textrm{Sgr~A}^*$ using the braneworld wormhole considered here. 

In the case of braneworld wormhole, the photon trajectories can be derived using identical procedures and there will be conserved energy $E$ as well as angular momentum $L$. Since the geometry is static and spherically symmetric, the Carter constant can be set to zero and hence, from \ref{eqalpha} and \ref{eqbeta}, it follows that, $\alpha^{2}+\beta^{2}=(L/E)^{2}$, i.e., the shadow in the Celestial coordinates will be a circle. However, the values of $L$ and $E$ depends on the details of the background geometry through the photon sphere and hence in this case depends on the ratio $p\equiv (\kappa/\lambda)$. Following which, we have plotted the shadow for various choices of the parameter $p$. As evident from \ref{Fig1-1}, as $p$ increases, the shadow radius decreases. This can be compared to the case of the ultra-compact object on the brane, considered in the previous section, for which as the parameter $q$ becomes more and more positive, the shadow radius decreases. Thus larger values of $p$ for a braneworld wormhole is in direct correspondence with positive tidal charge parameter $q$ for the ultra-compact object on the brane, as long as the shadow structure of these black hole alternatives are considered. This completes the theoretical study of the wormhole shadow, following which we now consider implications of the same from the observation of $\textrm{Sgr~A}^*$.

\section{Observables and implications from the shadow of $\textrm{Sgr~A}^*$}\label{Sec4}

With all the theoretical tools at hand, in this section, we compare the theoretically derived properties of the shadow in the higher dimensional scenario, with the observed image of the $\textrm{Sgr~A}^*$. This enables us in establishing possible constraints on the nature and properties of the extra dimension, in particular, whether one can assess if the presence of such an extra dimension is even consistent. For this purpose, we use the observables defined in \cite{Banerjee:2019nnj} and re-expressed below for clarity. First of all, we have the angular diameter $\Theta$ of the shadow, which is the primary observable, with the definition,
\begin{align}
\Theta=\frac{GM}{c^2}\left(\frac{\Delta\beta}{D}\right)~,
\label{Eq1}
\end{align}
where, $M$ is the mass of the object, $D$ is the distance of $\textrm{Sgr~A}^*$ from Earth and $\Delta \beta$ corresponds to the maximum length of the shadow along the $\beta$ direction in the $(\alpha,\beta)$ plane, in a direction orthogonal to $\alpha$. As evident, through $\Delta \beta$, the angular diameter depends on $(M,a,q)$, in the case of the ultra compact object on the brane, and on $(M,p\equiv \kappa/\lambda)$ for the wormhole on the brane. Alike the angular diameter, we can define another observable, namely, the axis ratio, which is given by,
\begin{align}
\Delta A=\frac{\Delta\beta}{\Delta\alpha}.
\label{Eq2}
\end{align}
where, alike $\Delta \beta$, $\Delta \alpha$ corresponds to the maximum separation between two points on the shadow boundary along the $\alpha$ direction on the $(\alpha,\beta)$ plane, in a direction orthogonal to $\beta$. Finally, for the rotating case, one can define the deviation of the shadow from its distinct circular appearance for static and spherically symmetric spacetime as another observable, taking the form \cite{Bambi:2019tjh},
\begin{align}
\Delta C= \dfrac{1}{R_{\rm avg}}\sqrt{\dfrac{1}{2\pi}\int_{0}^{2\pi}d\phi\left\{ \ell(\phi)-R_{\rm avg}\right\}^{2}}
\label{Eq4-3}
\end{align}
where, the average radius $R_{\rm avg}$ is defined as,
\begin{align}
R_{\rm avg}=\sqrt{\dfrac{1}{2\pi}\int_{0}^{2\pi}d\phi\, \ell^{2}(\phi)}
\label{Eq4-4}
\end{align}
and $\ell(\phi)^{2}=\left\{\alpha(\phi)-\alpha_{i} \right\}^{2}+\beta^{2}(\phi)$ being the distance between the geometrical centre ($\alpha_i,\beta_i$) of the shadow and any point ($\alpha,\beta$) on the shadow surface. The geometrical centre is simply obtained by averaging $\alpha$ and $\beta$ over the area. Since both the observables presented in \ref{Eq2} and \ref{Eq4-3} depends on details of the shadow structure, they inevitably depend on the hairs of the metric and hence on the extra dimension. For the rotating ultra compact object on the brane all of these observables are non-trivial, however, for the wormhole configuration on the brane we have, $\Delta A=1$ and $\Delta C=0$, respectively. Thus presence of a non-zero $\Delta C$, as well as non-zero values of $(\Delta A-1)$ would signal the existence of a rotating central object. However, for $\textrm{Sgr~A}^*$, there are several estimates of the rotation parameter from various different avenues and they predict $(a/M)$ values as large as $0.95$ to very small, or, zero rotation. For example, the motion of S2 stars near the $\textrm{Sgr~A}^*$, determines the spin of $\textrm{Sgr~A}^*$ to be $a\lesssim 0.1$ \cite{Fragione:2020khu}, while the study of its radio spectrum \cite{Reynolds:2013rva,Moscibrodzka:2009gw,Shcherbakov:2010ki} yields $a\sim 0.9$ \cite{Moscibrodzka:2009gw} and $a\sim0.5$ \cite{Shcherbakov:2010ki}. Further, investigation of the X-ray light curve of $\textrm{Sgr~A}^*$ reveals that the object has $a=0.9959 \pm 0.0005$ \cite{2010MmSAI..81..319A}. Thus in accordance with the numerical simulations \cite{EventHorizonTelescope:2022urf,EventHorizonTelescope:2022xqj} and the previous spin estimates by other methods, we therefore use $|a|=0$, $|a|=0.5$ and $|a|=0.94$, throughout this discussion.  Moreover, the Event Horizon Telescope collaboration also provides no data for the observables $\Delta A$ and $\Delta C$ for the shadow of Sgr A* and hence the zero rotation case must also be studied.  

\subsection{Implication from the observation of the shadow of $\textrm{Sgr~A}^*$}

\begin{figure*}[]
\begin{center}
\subfloat[\label{Fig2a}The above figure shows the variation of the angular diameter in the $q-a$ plane assuming $M=4.261\times 10^6 M_\odot$ and $D=8.2467$ kpc.]{\includegraphics[scale=0.45]{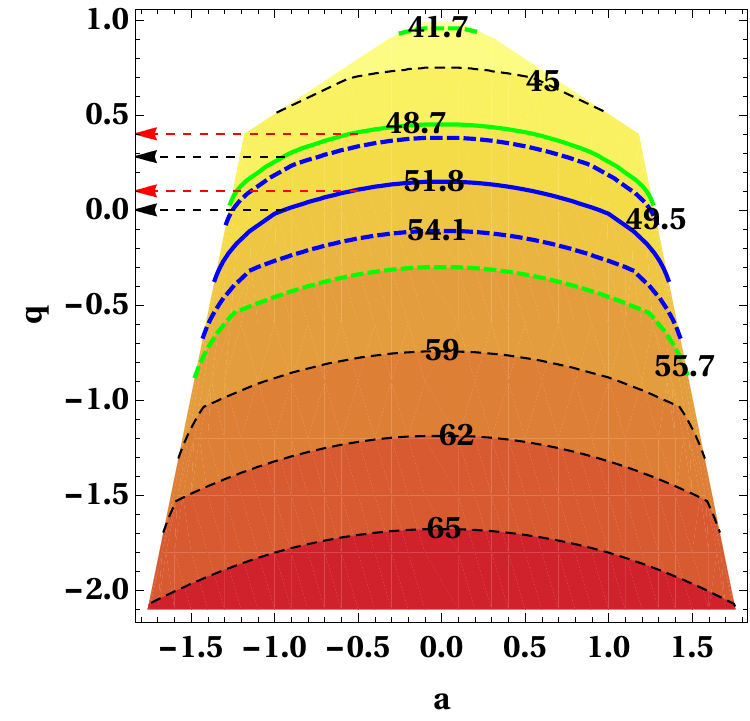}}
~~~~~~~~~~~
\subfloat[\label{Fig2b}The above figure depicts the variation of the angular diameter in the $q-a$ plane assuming $M=4.297\times 10^6 M_\odot$ and $D=8.277$ kpc.]{\includegraphics[scale=0.45]{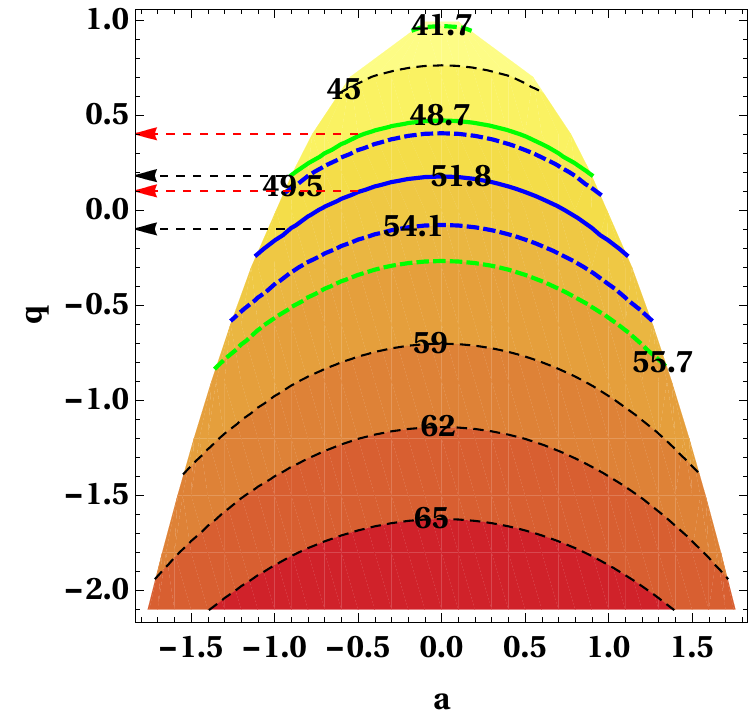}}
\\
\subfloat[\label{Fig2c} The dependence of the angular diameter on $q$ and $a$ is shown in the figure above. Here we assume $M=3.975\times 10^6 M_\odot$ and $D=7.959$ kpc to derive the angular diameter.]{\includegraphics[scale=0.45]{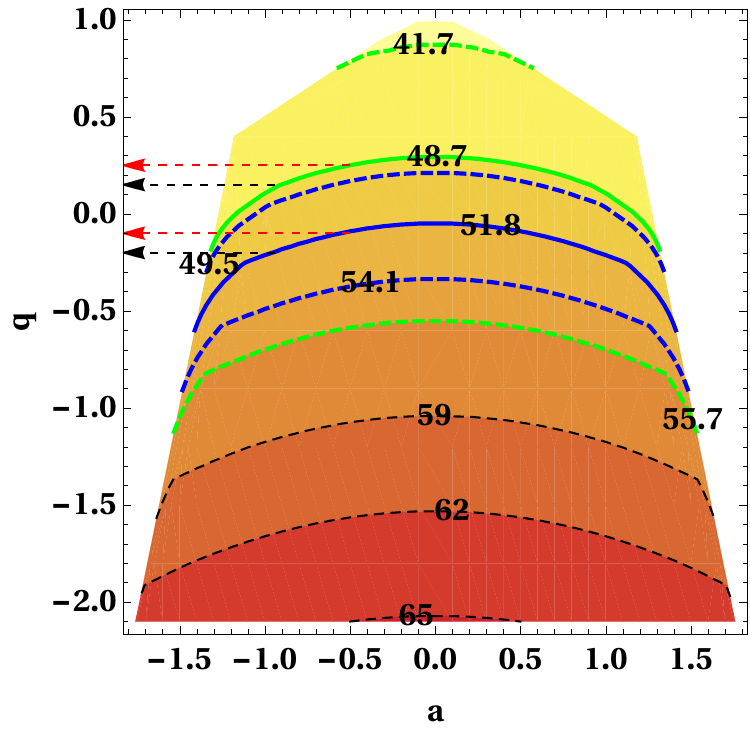}}
~~~~~~~~~~~
\subfloat[\label{Fig2d} The dependence of the angular diameter on $q$ and $a$ is shown in the figure above. Here the mass and distance are taken to be $M=3.951\times 10^6 M_\odot$ and $D=7.935$ kpc respectively.]{\includegraphics[scale=0.45]{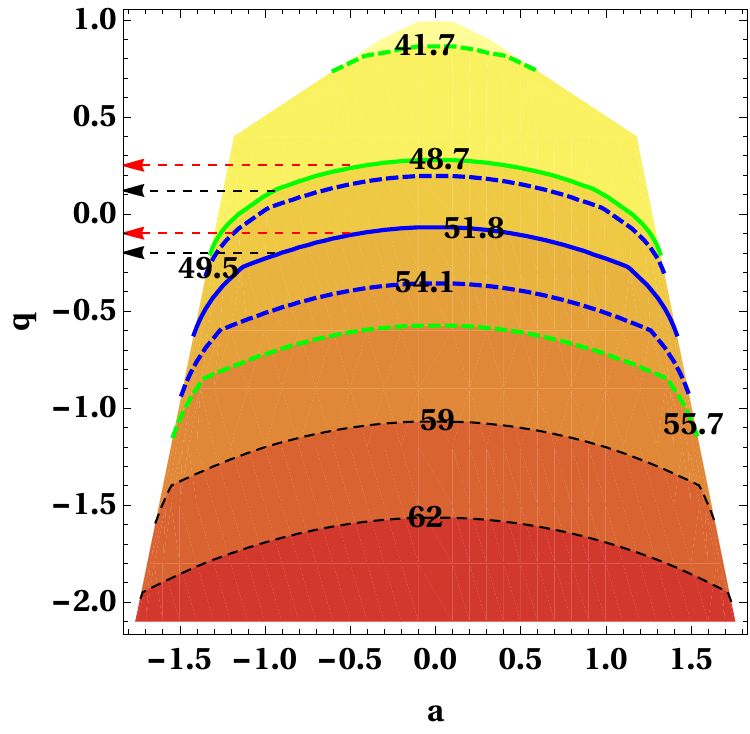}}
\hfill
\caption{The above figure illustrates the variation of the angular diameter with $q$ and $a$, with the inclination angle being $i=134^\circ$, while masses and distances are taken to be those reported by the Gravity collaboration (\ref{Fig2a} and \ref{Fig2b}) and the Keck team (\ref{Fig2c} and \ref{Fig2d}). This is compared with the observed image diameter and the shadow diameter, whose central values are respectively denoted by the blue and green solid lines while the 1-$\sigma$ contours are denoted by the blue and green dashed lines. Note that for negative values of the tidal charge parameter $q$, the rotation parameter can achieve the maximum value $1+|q|$, which is greater than unity. The same is reflected in the plots as well.}\label{Fig2}
\end{center}
\end{figure*}

Having detailed the observables, let us delve into the implications arising out of the results of the shadow measurements of $\textrm{Sgr~A}^*$. To start with, we note from \ref{Eq1} that in order to obtain the theoretical angular diameter of the shadow, one needs to provide the mass $M$ and the distance $D$ of the $\textrm{Sgr~A}^*$ from the Earth. According to the recent results of the Event Horizon Telescope collaboration \cite{EventHorizonTelescope:2022xnr,EventHorizonTelescope:2022vjs,EventHorizonTelescope:2022wok,EventHorizonTelescope:2022exc,EventHorizonTelescope:2022urf,EventHorizonTelescope:2022xqj}, the emission ring of $\textrm{Sgr~A}^*$ has an angular diameter of $51.8\pm 2.3 \mu as$ \cite{EventHorizonTelescope:2022xnr}, however, the angular diameter of the shadow itself turns out to be $48.7\pm 7 \mu as$ \cite{EventHorizonTelescope:2022xnr}. In order to be conservative, we consider both the angular diameters and try to see the consistency of these results with the presence of extra spatial dimension. The mass and the distance of the source, on the other hand, have been estimated by several groups. According to the results of the Gravity Collaboration, the mass and the distance of $\textrm{Sgr~A}^*$ are $M = (4.261 \pm 0.012) \times 10^6 M_\odot$ and $D=8246.7 \pm 9.3$ pc, respectively \cite{GRAVITY:2021xju,GRAVITY:2020gka}. When systematics due to optical aberrations are taken into account, the revised mass and distance turn out to be $4.297\pm 0.012\pm 0.040\times 10^6 M_\odot$ and $8277\pm 9 \pm 33$ pc, respectively. 
The Keck Collaboration, on the other hand, reports $M = (3.975 \pm 0.058 \pm 0.026) \times 10^6 M_\odot$ and $D = 7959 \pm 59 \pm 32 $ pc which are derived from fits that do not fix the redshift parameter at all \cite{Do:2019txf}. Fixing, the redshift parameter to unity, the mass and the distance, as measured by the Keck team turn out to be, $M = (3.951 \pm 0.047)\times 10^6 M_\odot$ and $D = 7935 \pm 50$ pc, respectively \cite{Do:2019txf}. The only other information required for the $\textrm{Sgr~A}^*$ is its inclination angle, and from models derived using extensive numerical simulation, reveals that the inclination angle of the source is constrained by $i<50^\circ$ \cite{EventHorizonTelescope:2022xnr}. 

\begin{figure*}[]
\begin{center}

\subfloat[\label{Fig3a} The dependence of the angular diameter of the shadow on the wormhole parameter $p$ is shown in the figure above. Here the mass and distance are taken to be $M=3.951\times 10^6 M_\odot$ and $D=7.935$ kpc, respectively, from the Keck collaboration. The theoretical estimation is being compared with the observational result from $\textrm{Sgr~A}^*$.]{\includegraphics[scale=0.55]{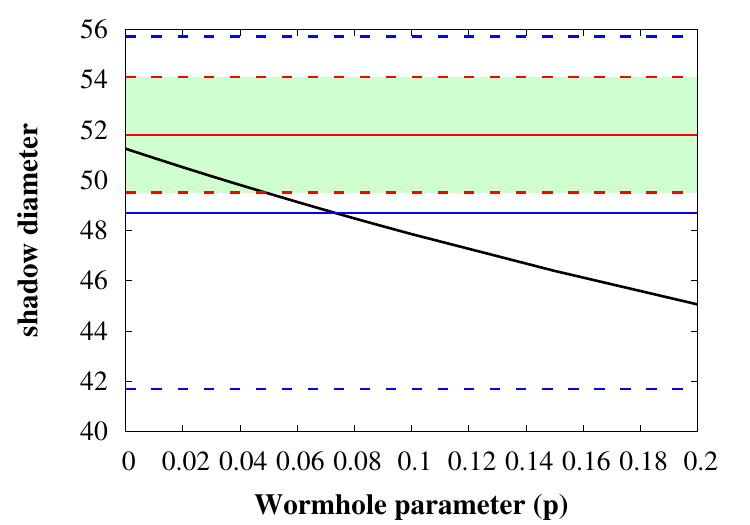}}
~~~~~~~~~~~~
\subfloat[\label{Fig3b} The dependence of the angular diameter of the shadow on the wormhole parameter $p$ is shown in the figure above. Here we assume $M=3.975\times 10^6 M_\odot$ and $D=7.959$ kpc, from the Keck collaboration. A comparison with the observational result from $\textrm{Sgr~A}^*$ has been presented.]
{\includegraphics[scale=0.55]{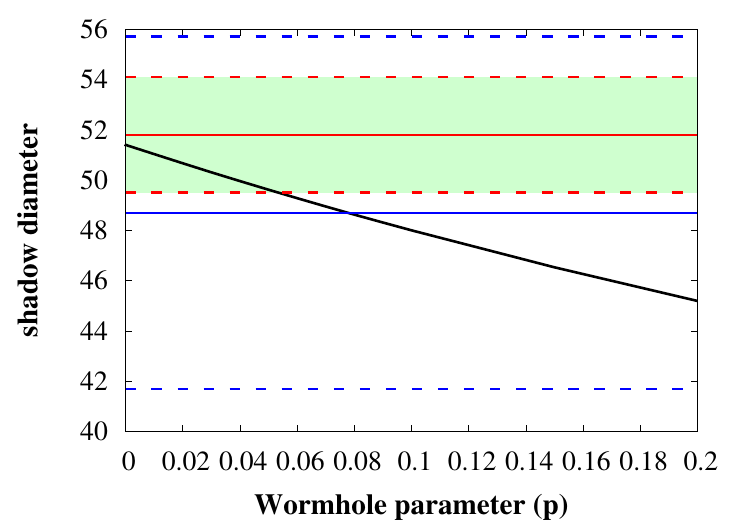}}
\\
\subfloat[\label{Fig3c}The above figure depicts the variation of the angular diameter of the shadow with the wormhole parameter $p$ assuming $M=4.261\times 10^6 M_\odot$ and $D=8.2467$ kpc, from the Gravity collaboration. Comparison with the observational result from $\textrm{Sgr~A}^*$ has also been presented.]{\includegraphics[scale=0.55]{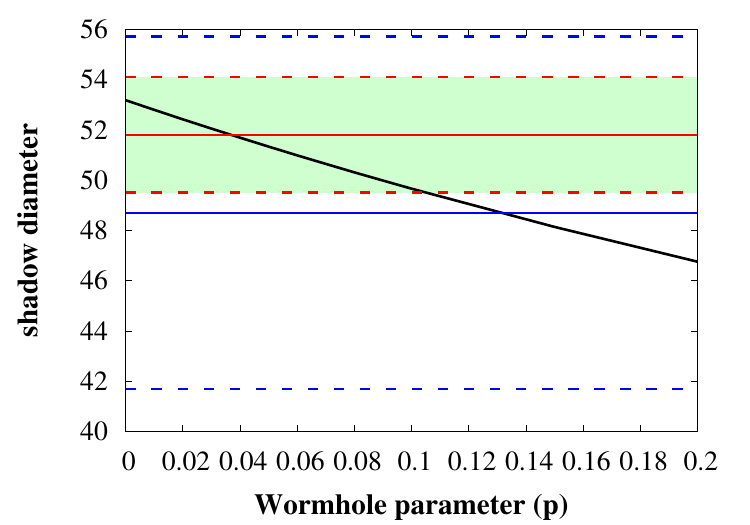}}
~~~~~~~~~~~~
\subfloat[\label{Fig3d}The above figure depicts the variation of the angular diameter with the wormhole parameter $p$ assuming $M=4.297\times 10^6 M_\odot$ and $D=8.277$ kpc, from the Gravity collaboration. The theoretical estimation of the shadow diameter can be compared with the observational result from $\textrm{Sgr~A}^*$.]{\includegraphics[scale=0.55]{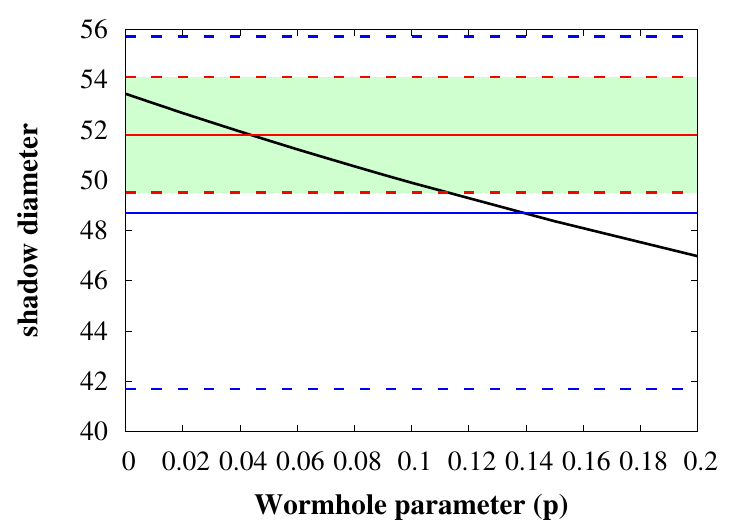}}

\caption{The above figures demonstrate the variation of the angular diameter with the wormhole parameter for various choices of the mass and the distance from the Keck and the Gravity collaboration. The red solid line corresponds to the central value of the observed image diameter while the red dashed lines represent the 1-$\sigma$ values. This entire region is shaded with green color. The shadow diameter is denoted by the blue lines where the solid line correspond to the central value while the dashed lines are associated with the 1-$\sigma$ limits.  In \ref{Fig3a} and \ref{Fig3b} the angular diameter is plotted assuming mass and distance reported by the Keck team while in \ref{Fig3c} and \ref{Fig3d} the angular diameter is evaluated assuming the mass and distance reported by the Gravity collaboration.
}
\label{Fig3}
\end{center}
\end{figure*}

In what follows, we present the theoretical computation of the angular diameter of the shadow of $\textrm{Sgr~A}^*$ using all of the above combinations of mass and distance in this work, while following \cite{2019A&A...625L..10G}, we use the inclination angle of the source to be $134^\circ$ (or equivalently $46^\circ$). The result of such a theoretical analysis, along with the observational data for the angular diameter of the shadow of $\textrm{Sgr~A}^*$ has been presented in \ref{Fig2} and \ref{Fig3}, for the ultra compact object on the brane and the wormhole on the brane, respectively. We first summarise our findings for the rotating compact object on the brane, through the constraints on the tidal charge $q$, from the comparison between the theoretical and the observed angular diameter of the shadow of $\textrm{Sgr~A}^*$, below:

\begin{itemize}

\item From \ref{Fig2a} and \ref{Fig2b}, where the theoretical angular diameter is obtained by using the mass and the distance reported by the Gravity collaboration \cite{GRAVITY:2021xju,GRAVITY:2020gka}. It is clear that the central value of the observed image diameter (51.8 $\mu as$, denoted by the blue solid line) as well as the observed shadow diameter (48.7 $\mu as$, denoted by Green solid line) can be best explained by non-zero value of the tidal charge, favouring the higher dimensional scenario. 

\begin{enumerate}

\item For \ref{Fig2a}, from the observed image diameter, we observe that, this corresponds to $q\simeq 0.1$ (for $|a|=0.5$ denoted by red dashed arrow), $q\simeq 0$ (for $|a|=0.94$, denoted by black dashed arrow) and $q\simeq 0.2$ (for $|a|=0$). Thus non-zero values of q, and hence presence of extra dimension is favoured by the observed image diameter. On the other hand, for observed shadow diameter, in \ref{Fig2a}, the $q$ values correspond to $q\simeq0.4$ (for $|a|=0.5$), $q\simeq0.3$ (for $|a|=0.94$) and $q\simeq 0.5$ (for $|a|=0$). The results mentioned above are for the central estimations of the mass and the distance of the $\textrm{Sgr~A}^*$. If we consider the uncertainties associated with these measurements as well, then the value of the tidal charge parameter will shift by an amount $\Delta q \sim 0.05$, well within the 1-$\sigma$ confidence contours.

\item While for \ref{Fig2b}, the observed image diameter is associated with $q\simeq 0.1$ (for $|a|= 0.5$), $q\simeq -0.1$ (for $|a|=0.94$) and $q\simeq 0.2$ (for $|a|=0$). We further note that when 1-$\sigma$ interval is considered in the image diameter (denoted by the blue dashed lines), both positive and negative values of $q$ are allowed. For the observed shadow diameter, these correspond to $q\simeq 0.4$ (for $|a|=0.5$), $q\simeq0.2$ (for $|a|=0.94$) and $q\simeq 0.5$ (for $|a|=0$). However, negative values of $q$ are included when 1-$\sigma$ contours are considered, which are denoted by the green dashed lines. Here also the inclusion of the uncertainties in the mass and distance measurements of the $\textrm{Sgr~A}^*$, yields the following uncertainty in the value of the tidal charge, $\Delta q\sim 0.16$, again within the 1-$\sigma$ contours.

\end{enumerate}

\item We next use the mass and the distance measurements reported by the Keck team, in order to obtain the theoretical angular diameter. Here also the central value of the observed image diameter (51.8 $\mu as$, denoted by the blue solid line) as well as the observed shadow diameter (48.7 $\mu as$, denoted by Green solid line) can be best explained by non-zero and negative values of the tidal charge, again favouring the higher dimensional scenario. 

\begin{enumerate}

\item In this case there is not much variation of the central q values. In particular, from \ref{Fig2c} and \ref{Fig2d} it is clear that the central value of the observed image diameter (blue solid line) can be best explained by $q\simeq-0.1$ (corresponding to $|a|=0.5$ denoted by red dashed arrow), $q\simeq-0.2$ (corresponding to $|a|=0.94$, denoted by black dashed arrow) and $q\simeq-0.04$ for $a=0$. The 1-$\sigma$ interval (denoted by the blue dashed lines) however includes both positive and negative values of $q$ and also includes the uncertainty in the tidal charge due to uncertainties in the mass and distance measurements of the $\textrm{Sgr~A}^*$, as $\Delta q\sim 0.34$.

\item Further, when we aim to reproduce the observed shadow diameter, for which the central value of $48.7\mu as$ (denoted by the green solid line) can be best explained by $q\simeq 0.25$ (when $|a|=0.5$), $q\simeq 0.1$ (when $|a|\simeq 0.94$) and $q\simeq 0.3$ for zero rotation. Once again, if the 1-$\sigma$ interval is considered (green dashed lines) both positive and negative values of the tidal charge are included, as well as the inclusion of the mass and the distance uncertainties of the $\textrm{Sgr~A}^*$ (yielding $(\Delta q/q)\sim 0.2$). 

\end{enumerate}

\end{itemize}

The above discussion reveals that in order to explain the observed shadow diameter, non-zero values of $q$ are favoured, for the observed image diameter, negative values of $q$ are favoured when mass and distance reported by the Keck team is used, while the positive values of the tidal charge better explains the observed image diameter with the mass and distance reported by the Gravity collaboration. The same holds true for the observed shadow diameter as well. Such a scenario, involving non-zero tidal charge, can be accommodated only in the context of the braneworld, since in the case of \gr\, the tidal charge must vanish, as astrophysical objects cannot have electric charge.  Also the unique provision of negative/positive tidal charge is realized in the higher dimensional framework alone and leads to an expansion/contraction of the angular diameter, which in turn serves to explain the observed image/shadow diameter better than the Kerr black hole scenario in \gr. Therefore, the observed shadow of $\textrm{Sgr~A}^*$ seems to exhibit a preference towards the braneworld scenario compared to \gr. Interestingly, the earlier observation of the shadow of M87* also indicates towards a similar conclusion \cite{Banerjee:2019nnj}.

So far we have discussed the possibility of $\textrm{Sgr~A}^*$ being a ultra compact object on the brane, which we found to be consistent with the observed image, as well as the shadow radius. Since, the spin of $\textrm{Sgr~A}^*$ is not well-measured and it can as well be zero, we would also like to explore the other possibility, where, $\textrm{Sgr~A}^*$ corresponds to a wormhole spacetime on the brane. In which case, the angular diameter will depend on the wormhole parameter $p\equiv (\kappa/\lambda)$, through $\Delta \beta$, the diameter of the shadow casted by the wormhole. In this case as well, we use the masses and distances measured by both the Keck team and the Gravity collaboration and determine the angular diameter, which we have presented in \ref{Fig3}. The following results can be obtained by comparing them with the observed angular diameter of the shadow:
 
\begin{itemize}

\item When the masses and distances, as estimated by the Keck team, are used we obtain \ref{Fig3a} and \ref{Fig3b}, respectively. As evident from both these figures, the central value of the shadow diameter can be best explained by a non-zero values of the wormhole parameter $p$ (blue solid line). Similarly, the following range of the wormhole parameter, $0\lesssim p \lesssim 0.06$ are allowed within  1-$\sigma$ confidence interval when the image diameter is considered (green shaded region). The same remains true even after the introduction of the mass and distance uncertainties, yielding $\Delta p \sim 0.02$.

\item In a similar footing, in \ref{Fig3c} and \ref{Fig3d} we plot the angular diameter of the shadow of the wormhole using the masses and the distances measured by the Gravity collaboration. From the two figures it is clear that a non-zero value of the wormhole parameter $p$ is required to reproduce the central value of both the image and the shadow diameter (red and blue solid line respectively). This holds true for the 1-$\sigma$ confidence interval as well. The introduction of the mass and distance uncertainties, yields $\Delta p \sim 0.06$, well within the 1-$\sigma$ confidence interval.

\end{itemize}
Thus we conclude that overall a non-zero values of the wormhole parameter $p$, inherited from the extra spatial dimension, explains the shadow of $\textrm{Sgr~A}^*$, better than the Schwarzschild scenario in \gr.

In this context, we should mention that we have bypassed the criticisms of testing alternative gravity models using the data from the event horizon telescope, as pointed out in \cite{Gralla:2020pra}, by considering the shadow diameter, as well as the image diameter. Of course, using any one of these observables will result into a dependence of the results on the details of the accretion physics. However, if both of these observables show an identical behaviour, i.e., they prefer a non-zero value of $q$, then it must arise from the fundamental physics and not from making wrong assumptions regarding the accretion model. As both of these diameters are affected equally by any errors made to the understanding of the Physics of accretion. Moreover, for robustness, we have also considered the results from both the Keck and the Gravity collaboration, as well as the mass and distance uncertainties associated thereof, even then the signatures of non-zero $q$ persisted. Therefore, such a generic feature cannot be arising from some erroneous assumption about the accretion model and hence the result must be taken seriously.

We should also point that, in order to be absolutely certain about the claim made in this work, we should have performed a simulation involving magneto-hydrodynamics of the plasma accreting onto the black hole or, the wormhole on the brane (henceforth as BraneMHD). However, such an analysis is not known in the braneworld scenario and is indeed an interesting future direction of exploration. In absence of such BraneMHD simulations, we have exploited all possible avenues available to us, e.g., we have not only considered the image diameter, but also the shadow diameter as reported by the Event Horizon Telescope collaboration. Besides, we have used the mass measurements from both the Keck team and the Gravity collaboration for $\textrm{Sgr~A}^*$. Most importantly, all of these different data suggest existence of non-zero $q$ or, non-zero wormhole parameter. This is the most important claim of this work, i.e., such a signature of higher dimensional scenario seems to be generic in the shadow of $\textrm{Sgr~A}^*$ and M87*, irrespective of the shadow or, image diameter and also independent of the mass measurements. This gives us hope that inclusion of BraneMHD in our analysis will further vote in favour of extra dimensional models, which we wish to come back to in future.

\section{Concluding Remarks}\label{Sec5}

In conclusion, we note that the braneworld scenario, fits the observed shadow and the image diameter of the $\textrm{Sgr~A}^*$ better, in comparison to the corresponding situation in \gr. To arrive at this result, we have considered two possible scenarios in the context of extra spatial dimension, namely the ultra compact rotating object on the brane and a wormhole on the brane. The ultra compact object inherits a tidal charge $q$ as the hair from extra dimension, through the bulk Weyl tensor, and relates to the size of the extra dimension. It turns out that, irrespective of the mass and the distance measurements from Gravity and Keck collaborations, the observed shadow and image diameter of the $\textrm{Sgr~A}^*$ always predicts a non-zero value of the tidal charge parameter $q$ within the allowed range of the angular diameter. In addition, as demonstrated above, the central values of the observed image and the shadow diameter explicitly provides $q\neq 0$ for both the Keck and the Gravity measurements of the mass and the distance of $\textrm{Sgr~A}^*$. 

Since the dimensionless spin parameter of  $\textrm{Sgr~A}^*$ is not-so-well measured, there is a possibility that it can be as low as zero as well. This prompts us to study if the central object of our galaxy, i.e., $\textrm{Sgr~A}^*$ can be a womhole. Even though wormholes, in general, require exotic matter to remain stable, in the presence of extra dimension such is not the case. Therefore, the wormhole solution considered here does not require exotic matter on the brane. To our surprise, it turns out that the observed shadow diameter predicts a non-zero value of $p$ (zero value would denote the Schwarzschild black hole), for the mass and the distance measurements from both the Keck and the Gravity observations. In particular, non-zero values of $p$ are also consistent with the 1-$\sigma$ confidence interval of the observed image diameter as well. 

Both of these results for $\textrm{Sgr~A}^*$, when coupled with the result of M87* \cite{Banerjee:2019nnj}, places the braneworld scenario to a more favourable position than general relativity. As demonstrated above, the observed shadow and the image diameter always predicts a non-zero value of the hair inherited from the extra spatial dimension, and holds true for the mass and the distance measurements of $\textrm{Sgr~A}^*$ from both the Keck and the Gravity collaborations. Though not conclusive, the above analysis provides a tantalizing avenue to search for extra spatial dimesions. Hopefully with more shadow observations pouring in, a joint analysis will reduce the error and provide a more conclusive hint towards extra spatial dimension. This result is also consistent with the outcome from other independent astrophysical observations. For example, most models explaining the observed quasi-periodic oscillations in black holes, show a preference towards a small but non-trivial negative tidal charge \cite{Banerjee:2021aln} and similar conclusions can be obtained from optical observations of a sample of eighty Palomar Green quasars \cite{Banerjee:2017hzw,Banerjee:2019sae} as well. It is intriguing that different observations with completely different samples of black holes, indicate towards a consistent result --- extra spatial dimensions exist. It is worth exploring other avenues, like gravitational waves and also wait for more observations associated with black hole shadow to appear in the near future, in order to have a conclusive evidence on extra spatial dimensions.    

\section*{Acknowledgements}

Research of S.C. is funded by the INSPIRE Faculty fellowship from DST, Government of India (Reg. No. DST/INSPIRE/04/2018/000893) and by the Start- Up Research Grant from SERB, DST, Government of India (Reg. No. SRG/2020/000409). S.C. further thanks the Albert-Einstein Institute, where a part of this work was carried out and the Max-Planck Society for providing the Max-Planck-India Mobility Grant.

\bibliography{Black_Hole_Shadow,Brane,KN-ED,regularBh,bardeen,SgrA,QPO,IB}

\bibliographystyle{./utphys1}
\end{document}